\documentclass[conference]{IEEEtran}
\usepackage[numbers,sort&compress]{natbib}
\usepackage{graphicx}
\usepackage{subcaption}
\usepackage{array}
\usepackage{booktabs}
\usepackage{url}
\begin{document}

%
\title{Multimodal Utterance-level Affect Analysis using Visual, Audio and Text Features}

\author{\IEEEauthorblockN{Didan Deng$^1$, Yuqian Zhou$^2$, Jimin Pi$^1$, Bertram E. Shi$^1$}
\IEEEauthorblockA{$^1$Department of Electronic and Computer Engineering, Hong Kong University of Science and Technology\\
$^2$IFP, Beckman, University of Illinois at Urbana-Champaign\\
$\{$ddeng,jpi,eebert$\}$@ust.hk, yuqian2@illinois.edu }
}

\maketitle
\thispagestyle{plain}
\pagestyle{plain}
\begin{abstract}
The integration of information across multiple modalities and across time is a promising way to enhance the emotion recognition performance of affective systems. Much previous work has focused on instantaneous emotion recognition. The 2018 One-Minute Gradual-Emotion Recognition (OMG-Emotion) challenge, which was held in conjunction with the IEEE World Congress on Computational Intelligence, encouraged participants to address long-term emotion recognition by integrating cues from multiple modalities, including facial expression, audio and language. Intuitively, a multi-modal inference network should be able to leverage information from each modality and their correlations to improve recognition over that achievable by a single modality network. We describe here a multi-modal neural architecture that integrates visual information over time using an LSTM, and combines it with utterance level audio and text cues to recognize human sentiment from multimodal clips. Our model outperforms the unimodal baseline, achieving the concordance correlation coefficients (CCC) of 0.400 on the arousal task, and 0.353 on the valence task.

\end{abstract}


%
\IEEEpeerreviewmaketitle

\section{Introduction}

Sentiment analysis or affective computing systems are designed to analyze human emotional states, and may benefit the development of human-computer interaction. The basic tasks include recognition of human sentiment using information from multiple modalities like facial expressions, body movement and gestures, speech and physiological signals. The labels for human sentiment are often either discrete categorical labels of six universal emotions (Disgust, Fear, Happiness, Surprise, Sadness, and Anger) \cite{ekman1971constants}, or continuous-valued annotations in the arousal and valence spaces \cite{thompson2011methods}. Previous research, therefore, has normally modeled the problem as either a classification\cite{zhou2017action} or a regression\cite{zhou2017pose} task, using deep models like the CNN\cite{khorrami2015deep}, or traditional approaches like the SVM or Regression Tree\cite{shan2009facial}. 

Further improvements in the performance and reliability of affective systems will rely on long-term contextual information modeling, and cross-modality analysis. Since emotions normally change gradually under the same context, analyzing long-term dependency of emotions will stabilize the overall predictions. Meanwhile, humans perceive others' emotional states by combining informatino across multiple modalities simultaneously. Combining different modalities will yield better emotion recognition with more human-like computational models \cite{morency2011towards}. These two aspects are explicitly emphasized in the 2018 IJCNN challenge "One-Minute Gradual-Emotion Recognition (OMG-Emotion)" \cite{barros2018omg}. In this challenge, long monologue videos with gradual emotional changes are selected from YouTube, and carefully annotated using both arousal/valence and emotional categories at the utterance-level. All the video clips contain visual, audio and transcript information. The performance of three unimodal recognition systems are provided as the baseline. 

In developing our multimodal system for sentiment analysis to address this challenge, we have been inspired by many previous works, such as that combining visual and audio features\cite{tzirakis2017end}, as well as speech content \cite{morency2011towards,poria2016fusing,zadeh2017tensor}. People have also combined physiological signals into emotion recognition systems \cite{ranganathan2016multimodal}. Methods of combining cues from each modality can be categorized into early or late fusion. For early fusion, features from different modalities are projected into the same joint feature space before being fed into the classifier \cite{rosas2013multimodal,poria2015towards}. For late fusion, classifications are made on each modality and their decisions or predictions are later merged together, e.g. by taking the mean or other linear combination \cite{cai2015convolutional,glodek2013kalman}. Some works\cite{kessous2010multimodal,poria2015deep} even implemented a hybrid fusion strategy to utilize both the advantages of late fusion and early fusion.

In this paper, we investigated the use of a number of feature extraction, classification and fusion methods. Our final trimodal method aggregates visual, audio and text features for a single-shot utterance-level sentiment regression using early fusion. To verify the effectiveness of multimodal fusion, we compared it with three unimodal methods. Our proposed multimodal approach outperformed the unimodal ones as well as the baseline methods, achieving validation set concordance correlation coefficients (CCC) of 0.400 on the arousal task, and 0.353 on the valence task.

\begin{figure*}[ht]
      \centering
      \includegraphics[width=0.8\linewidth]{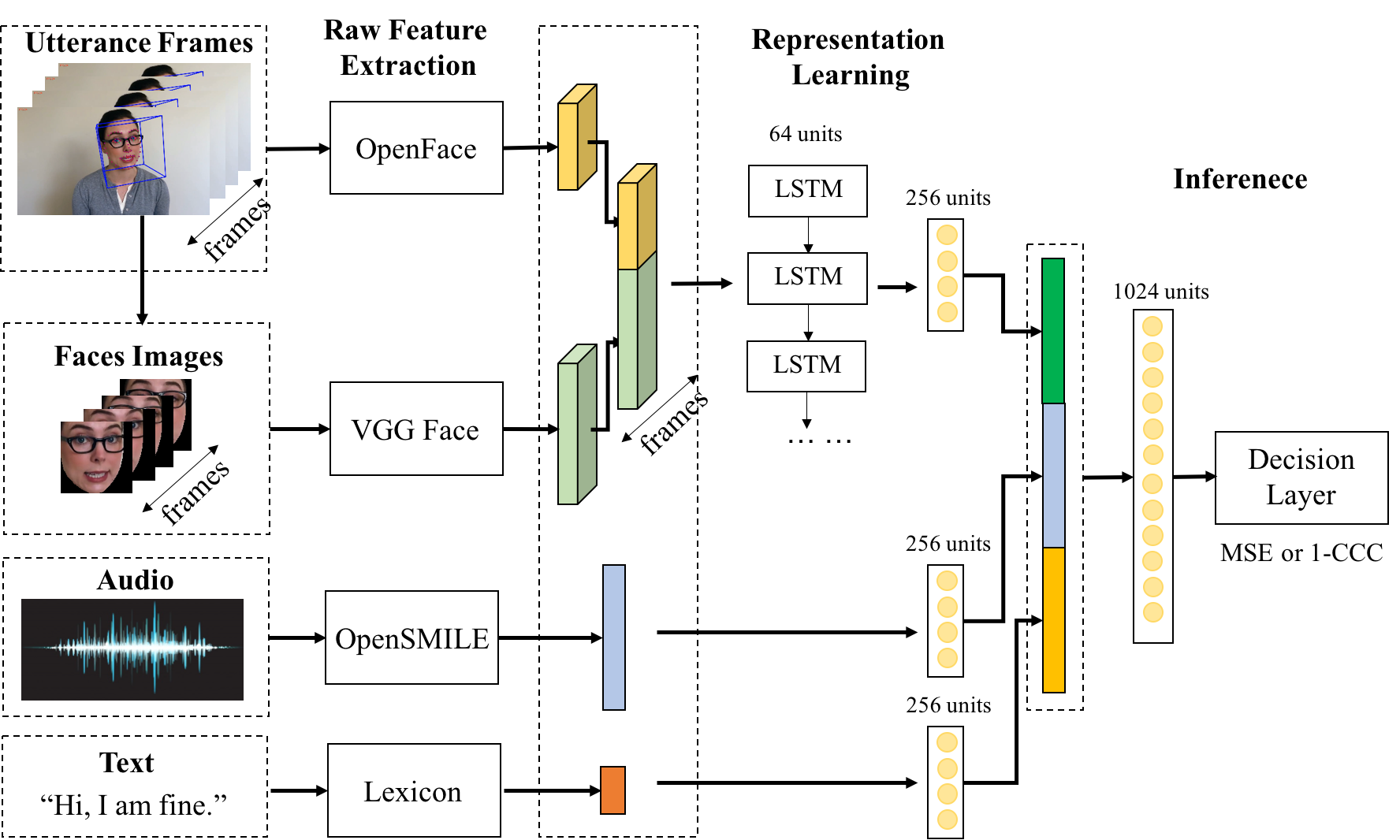}
      \caption{The architecture of the proposed model. The unimodal features are extracted separately and concatenated in an early fusion strategy. A two-layer fully-connected neural network is applied to estimate the estimate the arousal and valence of a single utterance.}
      \label{fig:structure}
\end{figure*}

\section{Methodology}
\subsection{Dataset and Metrics}

The OMG-Emotion Behavior Dataset\cite{barros2018omg} is a long-term multi-modal corpus for sentiment analysis. It is constructed by picking out the videos with emotion behaviors from Youtube videos using keywords like "monologues", "auditions" etc. Most videos in OMG dataset have standard resolution of 1280x720, and the main language is English. Utterances are then extracted from each video where there are high speech probability. The dataset is split into training, validation and testing set. There are 231 videos in the training set, 60 videos in the validation set, and 204 videos in the testing set. Thus the number of utterances are 2440, 617 and 2229 respectively.

Each utterance is annotated by arousal/valence value in dimensional space, as well as seven discrete emotion labels. Arousal is a continuous score ranging from 0 (calm) to 1 (excited), while valence is a continuous score ranging from -1 (negative) to +1 (positive).  

Two following metrics are used to evaluate the arousal/valence estimation over this dataset: MSE (mean squared error) and CCC (the concordance correlation coefficients). The CCC is defined as: 
\begin{equation}
\rho_{c} = \frac{2\rho \sigma_{Gnd} \sigma_{Pred}}{\sigma_{Gnd}^2 + \sigma_{Pred}^2 + (\mu_{Gnd} - \mu_{Pred})}
\end{equation}
where $\rho$ is the Correlation Coefficient between the predictions and groundtruth. $\mu_{Gnd}$ and $\mu_{Pred}$ denote the mean, and $\sigma_{Gnd}^2$ and $\sigma_{Pred}^2$ are the corresponding variance. 

\subsection{System Architecture}

Figure \ref{fig:structure} shows the architecture of our proposed model. Our deep neural network model consists of three parts: (1) the subnetworks for each single modality; (2) the early fusion layer which concatenates three unimodal representations together; and (3) the final decision layer that estimates the sentiment. 

\subsubsection{Visual Subnetwork}
Visual features consist of OpenFace \cite{baltruvsaitis2016openface} estimators on the whole frames, and VGG face representation \cite{parkhi2015deep} on facial regions. For OpenFace features, we use OpenFace toolkit to extract the estimated 68 facial landmarks in both 2D and 3D world coordinates, eye gaze direction vector in 3D, head pose, rigid head shape, and Facial Action Units intensity\cite{ekman1978facial} indicating the facial muscle movements. The detailed feature descriptions are seen in\cite{Tadas2018openface} Those visual descriptors are regarded as strong indicators of human emotions and sentiments \cite{ranganathan2016multimodal,soleymani2012multimodal}. For the VGG face representation, facial region in each frame is cropped and aligned using a 3D Constrained Local Model described in \cite{baltruvsaitis20123d}. We zero out the background according to the face contour indicated by the facial landmarks. Then, the cropped faces are resized to 224$\times$224$\times$3 and fed into a VGG Face model pretrained on a large face dataset. We take the 4096-dimensional feature vectors in the fc6 layer, and concatenate them with the visual features extracted by OpenFace. The total dimension of the concatenated features is 4805. 

The concatenated visual features from a single utterance are further fed into a LSTM layer with 64 hidden units followed by a dense layer with 256 hidden neurons for temporal modeling. Specifically, 20 frames are uniformly sampled from each utterance and fed into the network for training and testing. In the case of shorter length of utterance, we duplicated the last frame to fill the gap.

\subsubsection{Audio Subnetwork}\label{sec:audio}
Audio features are extracted using openSMILE toolkit\cite{eyben2010opensmile}, and we use the same feature set as suggested in the INTERSPEECH 2010 paralinguistics challenge\cite{schuller2010interspeech}. The set contains Mel Frequency Cepstral Coefficients (MFCCs), $\Delta$MFCC, loudness, pitch, jitter, etc.\cite{emobase2010}. These features describe the prosodic pattern of different speakers and are consistent signs of their affective states. For each utterance sample, We extract 1582 dimensional features from the audio signal. These audio features are then fed into a fully connected layer with 256 units.

\subsubsection{Text Subnetwork}
We use two opinion lexicons to analyze the patterns in language context. The first one is Bing Liu’s opinion Lexicon\cite{ding2008holistic} with 2006 positive words and 4783 negative words. The second one is MPQA Subjectivity Lexicon\cite{wilson2005recognizing} with 2718 positive words and 4913 negative words. For each utterance, we compute the frequency of positive and negative words according to the two lexicons, as well as the total word number in the whole utterance. For utterances without transcript, we replicate the transcript of the closest utterance in time. We also extract the word frequencies over the entire video, and assign them as features for all utterances in the same video. The total dimension of word feature is finally 10, including utterance-level and video-level word frequency from two lexicons and the total word counts. These text features are also fed into a fully connected layer with 256 units.

\subsubsection{Fusion and Decision Layers} \label{sec:declayer}
We combine cues from the three modalities using early fusion strategy. The aggregated feature vector is fully connected to a two-layer neural network with 1024 hidden units and a single output neuron, activated by sigmoid (for arousal task) or hyperbolic tangent function (for valence task). We first use MSE as the loss function for joint training, and apply $1-\rho_{c}$ loss for further refinement.

In comparison, we also design a late fusion strategy. In this case, we add a decision layer in each subnetwork and combine the 3 predictions using a linear regression trained by MSE.

\begin{table}[t]
\caption{The Ablation Test of Unimodal Models}
\label{tab:unimodal}
\begin{center}
\begin{tabular}{lllll}
\toprule
\multicolumn{1}{c}{\bf Models}  &\multicolumn{2}{c}{\bf Arousal} &\multicolumn{2}{c}{\bf Valence} \\
&(CCC) & (MSE) & (CCC) & (MSE) \\
\midrule
Visual(VGG-Face)        &0.109   & 0.047   &0.237  &0.110\\
Visual(OpenFace)          &0.046   &0.047  & 0.080 &0.122\\
Visual(Fused Feature)          &\bf0.175   &0.047  & \bf0.261 &0.122\\
\midrule
Audio(with LSTM)        &0.146   & 0.044   &0.154  &0.106\\
Audio(without LSTM)          &\bf0.273   &0.054  & \bf0.266 &0.108\\
\midrule
Text(Word Embedding)        &0.007   & 0.048   &0.098  &0.120\\
Text(Lexicon)          &\bf0.137   &0.044  & \bf0.259 &0.108\\
\bottomrule
\end{tabular}
\end{center}
\end{table}
\begin{table}[t]
\caption{The performance of two fusion strategies}
\label{tab:fusion}
\begin{center}
\begin{tabular}{lllll}
\toprule
\multicolumn{1}{c}{\bf Fusion Methods}  &\multicolumn{2}{c}{\bf Arousal} &\multicolumn{2}{c}{\bf Valence} \\
&(CCC) & (MSE) & (CCC) & (MSE) \\
\midrule
Late Fusion           &0.311   &0.046  & 0.280 &0.106\\
Early Fusion         &\bf0.386   & 0.054   &\bf0.305  &0.105\\
Early Fusion(Fine Tuned)         &\bf0.400   & 0.058   &\bf0.353  &0.136\\
\bottomrule
\end{tabular}
\end{center}
\end{table}

\section{Experiments}
We trained and evaluated the multimodal network on OMG dataset. The model was trained for at most 300 epochs. To prevent overfitting, we applied an early-stopping policy with 20 epochs patience, which means to stop training after the validation loss doesn't drop for 20 epochs, and we deployed dropout strategy with ratio 0.5 for each fully connected layer. The learning rate was $1e^{-2}$ for arousal task and $1e^{-3}$ for valence task.

\subsection{Unimodal Approach}
\begin{table*}[t]
\caption{The Performance on the Validation Partition}
\label{tab:final}
\begin{center}
\begin{tabular}{lllllllll}
\toprule
\multicolumn{1}{c}{\bf Model}  &\multicolumn{4}{c}{\bf Arousal} &\multicolumn{4}{c}{\bf Valence} \\
\toprule
 &\multicolumn{2}{c}{\bf Baseline}  &\multicolumn{2}{c}{\bf Ours} &\multicolumn{2}{c}{\bf Baseline}  &\multicolumn{2}{c}{\bf Ours}\\
  &\multicolumn{1}{c}{CCC}  &\multicolumn{1}{c}{MSE}  &\multicolumn{1}{c}{CCC}  &\multicolumn{1}{c}{MSE} &\multicolumn{1}{c}{CCC}  &\multicolumn{1}{c}{MSE} &\multicolumn{1}{c}{CCC}  &\multicolumn{1}{c}{MSE}\\
 
\midrule
Audio        &0.122   & 0.04   &\bf 0.273  &0.054  &0.049   &0.013  &\bf  0.266 &0.108\\
Video           &0.159   &0.05  & \bf 0.175 &0.047  &0.219   &0.15  &\bf  0.261 &0.122\\
Text           &0.003   &0.04  &\bf 0.137 &0.044  &0.068  &0.13  & \bf0.259  &0.108\\
Trimodal      &None   &None  &\bf 0.400 &0.058 &None   &None  & \bf 0.353 &0.136\\
\bottomrule
\end{tabular}
\end{center}
\end{table*}
We first evaluated the performance of model trained with single modality. For each unimodal model, the same decision layer introduced in Section \ref{sec:declayer} was deployed.

For visual unimodal model, we investigated the effectiveness of VGG-face and OpenFace features separately in an ablation test. The comparison results are shown in Table \ref{tab:unimodal}. Our results demonstrated that VGG-face features outperformed OpenFace features under the same model architecture. Better performance on both arousal and valence tasks were achieved when the two features are fused.

For the audio network, we focused on studying the importance of temporal modeling in utterance. We implemented another LSTM-based network for audio modality. Specifically, we divided each audio file into audio frames of 0.5 second length, and extracted openSMILE features for each single frame. Those features are then fed into a 64 cells LSTM layer followed by the decision layer. We compared this LSTM-based model with our audio unimodal model described in section \ref{sec:audio}. The results in Table \ref{tab:unimodal} show the model without LSTM performs better than the audio model with LSTM. The LSTM layer does not benefit the estimation.

For text modality, we compared the proposed word frequency statistical approach with models using pretrained word embeddings and LSTM layers in NLP(Natural Language Processing). We implemented the latter approach by using the 100 dimensional GloVe word vectors pretrained on English WikiPedia\cite{pennington2014glove} and a 64 cells LSTM layer in Text(LSTM) model. We compared the performance with text unimodal model using simple opinion lexicon features. The result is shown in Table \ref{tab:unimodal}. Surprisingly, simple lexicon features performed better. This results from the frequently occurring errors as being transcribed using Automatic Speech Recognition Tool in this dataset. The opinion lexicon features, however, mostly ignore these errors by only counting the words appearing in opinion lexicon.

\subsection{Multimodal Approach}

We trained the trimodal network by using the concatenated multimodal features. With respect to fusion strategies, We compared the early and late feature fusion strategies in Table \ref{tab:fusion}. The results demonstrated that learning benefits more from early fused representation. The performance is further improved by fine-tuning the system using $1-\rho_{c}$ loss. Table \ref{tab:final} showed the comparison of our unimodal or multimodal systems performances with the baseline results. The trimodal model has better performance than any of the unimodal models. 

\section{Conclusion}
In this paper, we propose a multimodal system that utilizes visual, audio and text features to perform a continuous affect prediction task in utterance level. Early feature fusion strategy is deployed and CCC loss is directly applied for network fine-tuning to boost the estimation performance. In the OMG dataset, both our unimodal or multimodal models outperform the baseline methods significantly. Our results shows that cross-modal information will greatly benefit the estimation of long-term affective states.



%

\bibliographystyle{ieeetr}
\bibliography{sample}

\end{document}